\documentclass[conference,9pt]{IEEEtran}
\usepackage{amsmath,amssymb}
\usepackage{graphicx}
\usepackage{cite}
\usepackage{url}
\usepackage{booktabs}
\usepackage{multirow}
\usepackage{tikz}
\usetikzlibrary{positioning,arrows.meta,fit}
\usepackage{xcolor}
\usepackage{enumitem}

\newcommand{\eg}{e.g.,\ }
\begin{document}

\title{SlotFill: Repairing Lossy SoC Communication Traces for Post-Silicon Validation}

\author{\IEEEauthorblockN{Tasnim Tabassum and Hao Zheng}\IEEEauthorblockA{University of South Florida, \{tasnimtabassum, haozheng\}@usf.edu}}

\maketitle
\thispagestyle{plain}
\pagestyle{plain}

\begin{abstract}
Post-silicon validation relies on traces of messages exchanged by SoC components, but trace buffers and monitors drop messages, often leaving no mark of what was lost. We present SlotFill, which puts each deleted message back in the right gap, with the right type and order. SlotFill gives every gap between the remaining messages a fixed number of slots, and a Transformer-based model fills each slot with a message or leaves it empty when nothing was deleted there, so the input never reveals how much was lost. The model also uses monitor fields (timestamps, transaction IDs, addresses) from surviving messages. On whole-program gem5 traces of 4- and 8-core systems (2.07 billion messages), SlotFill restores up to 97\% of deleted messages exactly at 15\% loss and up to 87\% at 50\% loss, and 76--83\% at 15\% loss when timestamps are coarse.
\end{abstract}


\section{Introduction}
Post-silicon validation is a large share of SoC validation cost, and it relies on traces of the messages that components exchange~\cite{mishra2017postsilicon}. These traces are incomplete because trace buffers are small, only a few links can be observed at once~\cite{ko2009restoration,cao2019selection}, and monitors that share one trace port must drop transactions when their queues overflow~\cite{cao2018monitors}. A dropped message is simply absent: at best the trace flags that an overflow happened, but not where or what was lost. Every analysis built on the trace then sees broken flows. Scenario reconstruction~\cite{cao2017trace} and flow mining~\cite{ahmed2024automodel, nadimi2025autoflows} assume that each request has its response and each flow is complete.

We study \emph{deleted-message recovery}: given a damaged trace, put each missing message back (1)~in the right gap between surviving messages, (2)~with the right message type, and (3)~in the right order inside the gap.
 Here the model must also decide \emph{whether} anything is missing at each position, in a trace where many flows run at the same time and interleave. The only prior work on this exact task is a master's thesis~\cite{fender2024}. It used BERT/BART models in a two-step find-then-fill pipeline and recovered 10.1\% (synthetic) and 33.7\% (gem5) of deleted messages.

We propose \textbf{SlotFill}. Its core is a \emph{slot layout}: every gap gets the same number $S$ of slots, and each slot holds either a message or \textsc{null} (empty). Recovery becomes a fill-in-the-blanks task with a fixed shape, which a standard Transformer can learn. Finding the gap, the type and the order is one decision per slot, not three separate steps. Because every gap looks the same, the input never reveals how many messages were lost. The layout does not fix \emph{how} the slots are filled, so we test two common ways: left to right, one slot at a time, and in parallel, by masked diffusion~\cite{sahoo2024mdlm}. The model also reads the fields that hardware monitors record: timestamps, transaction IDs, addresses and request/response direction~\cite{arm2023axi,arm2017coresight,cao2018monitors}.

This paper sets out to (1)~define and solve deleted-message recovery without placeholders: we design the slot layout, whose length leaks no deletion count, a simple bound for the number of slots, and metrics that count only real messages (exact, tie-aware, flow-relative); (2)~utilize the monitor fields, taken \emph{only} from surviving messages and checked by a leak audit, and measure their impact as less field information is available; (3)~find how small a model suffices: on whole gem5~\cite{binkert2011gem5,lowepower2020gem5} SE and FS runs of 4- and 8-core systems (2.07 billion messages), 5M-parameter models restore over 90\% of deleted messages exactly at 15\% loss with either decoder (left to right 1.6--4.7 points better), and larger models add nothing; and (4)~show that repair helps trace checking: against an AutoModel~\cite{ahmed2024automodel} checker, it removes most false alarms on surviving messages and brings accepted events back to 93--100\% of the clean trace, while barely changing how many injected bugs are rejected.

We restore the gap, type and order of each lost message, not its fields, so that checks on a damaged trace behave as on a complete one.

\section{Background and Related Work}
\textbf{Post-silicon observability.} Trace-signal selection chooses which signals to record so that others can be restored~\cite{ko2009restoration}. For SoC protocols, Cao et al.\ reconstruct flow scenarios from partial traces~\cite{cao2017trace}, build transaction monitors that compress link activity into messages~\cite{cao2018monitors}, and select which links to monitor under a bandwidth budget~\cite{cao2019selection}. They show that fields such as transaction tags and sequence IDs strongly reduce ambiguity~\cite{cao2017trace}. These works decide what to record; we repair what was lost anyway.

\textbf{Mining communication traces.} Flow and model mining infer protocols from traces, for example by constraint solving~\cite{ahmed2024automodel} or causality graphs~\cite{nadimi2025autoflows}. They assume complete traces, so recovery is a natural pre-processing step for them.

\textbf{Repairing sequences.} In process mining, missing events in logs are restored with the help of a given process model~\cite{wang2013missing}; SoC specifications are usually incomplete, so we do not assume one. Time-series imputation, including diffusion-based imputation~\cite{tashiro2021csdi}, knows where values are missing. Fender's~\cite{fender2024} is the closest work: its find-then-fill cascade makes errors in step one that step two cannot fix, and its sequence-to-sequence variant rewrote the whole trace and produced outputs with more errors (25\% token error) than the 15\%-damaged input.

\textbf{Filling sequences.} Transformers~\cite{vaswani2017attention} generate text left to right; masked diffusion models~\cite{austin2021d3pm,sahoo2024mdlm} fill masked tokens in parallel, most confident first~\cite{ghazvininejad2019maskpredict,chang2022maskgit}, and now scale to 7--8B parameters~\cite{nie2025llada,ye2025dream}. Insertion models add tokens between existing ones: one at a time~\cite{stern2019insertion}, by first predicting how many each gap needs~\cite{gu2019levenshtein}, or by growing blanks to any length~\cite{shen2020blank}. SlotFill instead gives every gap the same fixed number of slots, so the input hides how much was lost, and one label per slot says where, what and in which order. Our time features build on rotary embeddings~\cite{su2024roformer} and event-time encodings~\cite{zuo2020thp,kazemi2019time2vec}.

\section{Problem Statement}
A trace is a sequence $\rho=(m_1,\dots,m_n)$. Each message has a \emph{type} \texttt{src:dst:cmd:req|resp} (\eg \texttt{cpu0:dcache0:ReadReq:req}) and the fields a monitor records: tick, transaction ID (txid), cache-line address, direction, and a no-reply flag for requests that get no response (\eg \texttt{CleanEvict}). Loss removes a set $D$ of messages at rate $r$. Each loss event randomly removes a set of consecutive messages at a random position; events that land next to each other merge, so one gap can lose many messages (Sec.~\ref{sec:numslots}). The damaged trace $\tilde\rho$ keeps the other $K$ messages, in their original order. Deleted messages can only sit between them, so $\tilde\rho$ has $K{+}1$ \emph{gaps}: gap~0 before the first kept message, gap~$k$ between kept messages $k$ and $k{+}1$, and gap~$K$ after the last one. Most gaps are empty. Given a damaged trace $\tilde\rho$, the task is to find a repaired trace $\hat\rho$ close to $\rho$: as many deleted messages as possible restored in the right gap, type and order, with few inserted messages that were never there (Sec.~\ref{sec:metrics}).

\section{Methodology}
\begin{figure}[t]
\centering
\begin{tikzpicture}[font=\scriptsize, x=1.25cm, >=Stealth,
  k/.style={draw, fill=blue!12, minimum width=7.2mm, minimum height=4mm, inner sep=1pt},
  s/.style={draw, dashed, fill=gray!8, minimum width=3.8mm, minimum height=4mm, inner sep=0pt},
  f/.style={draw, fill=orange!25, minimum width=3.8mm, minimum height=4mm, inner sep=0pt},
  hw/.style={draw, rounded corners=1pt, fill=green!12, minimum width=8mm, minimum height=5mm, inner sep=1pt},
  mon/.style={draw, fill=yellow!35, minimum width=6mm, minimum height=3.6mm, inner sep=1pt, font=\tiny}]
\def\gapA{0.16,0.52,0.88} \def\gapB{1.94,2.30,2.66} \def\gapC{3.72,4.08,4.44}
\node[hw] (cpu) at (0.32,2.62) {CPU};
\node[hw] (l1)  at (2.22,2.62) {L1 \$};
\draw[<->, thick] (cpu) -- (l1);
\node[mon] (m) at (1.27,2.62) {monitor};
\node[font=\tiny, align=center, fill=yellow!12, draw, inner sep=1.5pt] (rec) at (1.27,2.1) {message\,$\cdot$\,tick\,$\cdot$\,txid\,$\cdot$\,addr\,$\cdot$\,dir\,$\cdot$\,type};
\draw[->] (m.south) -- (rec.north);
\foreach \i in {0,...,17} {\fill[blue!40] (2.95+0.095*\i,2.0) rectangle ++(0.07,0.2);}
\draw[->] (rec.east) -- (2.91,2.1);
\node[font=\tiny] at (3.84,2.42) {whole-run ROI};
\draw[densely dashed, thick, orange!80!black] (3.50,1.94) rectangle (4.18,2.26);
\node[font=\tiny, orange!50!black] at (3.84,1.8) {random window};
\node[draw, fill=red!8, inner sep=1.5pt, font=\tiny, align=center] (del) at (5.12,2.1) {random\\deletion};
\draw[->] (4.64,2.1) -- (del.west);
\draw[->, thick, violet!70!black] (del.south) -- node[right, font=\tiny, inner sep=5pt] {repair} (5.12,1.6);
\draw[rounded corners=3pt, violet!70!black, fill=violet!4] (-0.12,-1.42) rectangle (5.72,1.6);
\node[font=\tiny, fill=violet!10, draw=violet!70!black, rounded corners=1pt, inner sep=1.5pt] at (2.8,-1.42) {\textbf{test:} SlotFill decoder, left to right (AR) or parallel (MDLM)};
\node[anchor=west] at (-0.05,1.42) {\textbf{(a)} damaged trace: only kept messages, with fields};
\node[k] at (1.41,1.0) {$k_1$}; \node[k] at (3.19,1.0) {$k_2$}; \node[k] at (4.97,1.0) {$k_3$}; \node at (5.45,1.0) {\dots};
\node[font=\tiny] at (1.41,0.72) {tick 10}; \node[font=\tiny] at (3.19,0.72) {tick 19}; \node[font=\tiny] at (4.97,0.72) {tick 19};
\node[anchor=west] at (-0.05,0.36) {\textbf{(b)} slot layout: $S$ slots per gap (here $S{=}3$), all unknown};
\foreach \x in \gapA {\node[s] at (\x,0) {\tiny ?};} \foreach \x in \gapB {\node[s] at (\x,0) {\tiny ?};} \foreach \x in \gapC {\node[s] at (\x,0) {\tiny ?};}
\node[k] at (1.41,0) {$k_1$}; \node[k] at (3.19,0) {$k_2$}; \node[k] at (4.97,0) {$k_3$}; \node at (5.45,0) {\dots};
\node[anchor=west] at (-0.05,-0.64) {\textbf{(c)} repaired: each slot becomes a message or $\varnothing$};
\foreach \x in \gapA {\node[s] at (\x,-1.0) {$\varnothing$};} \foreach \x in \gapC {\node[s] at (\x,-1.0) {$\varnothing$};}
\node[f] at (1.94,-1.0) {$x$}; \node[f] at (2.30,-1.0) {$y$}; \node[s] at (2.66,-1.0) {$\varnothing$};
\node[k] at (1.41,-1.0) {$k_1$}; \node[k] at (3.19,-1.0) {$k_2$}; \node[k] at (4.97,-1.0) {$k_3$}; \node at (5.45,-1.0) {\dots};
\end{tikzpicture}
\vspace{-1mm}
\caption{SlotFill overview. A monitor on a link records every message with its fields (tick, txid, address, direction, type). We take a random window from a whole-run ROI trace and delete messages at random to simulate trace loss. SlotFill then repairs the window: (a) only kept messages survive, but their fields give hints: 9 cycles passed between $k_1$ and $k_2$. (b) Every gap gets $S$ slots, so the length depends only on the number of kept messages. (c) The decoder (left to right or parallel) fills each slot with a message ($x$, $y$) or $\varnothing$ (empty).}
\label{fig:overview}
\end{figure}
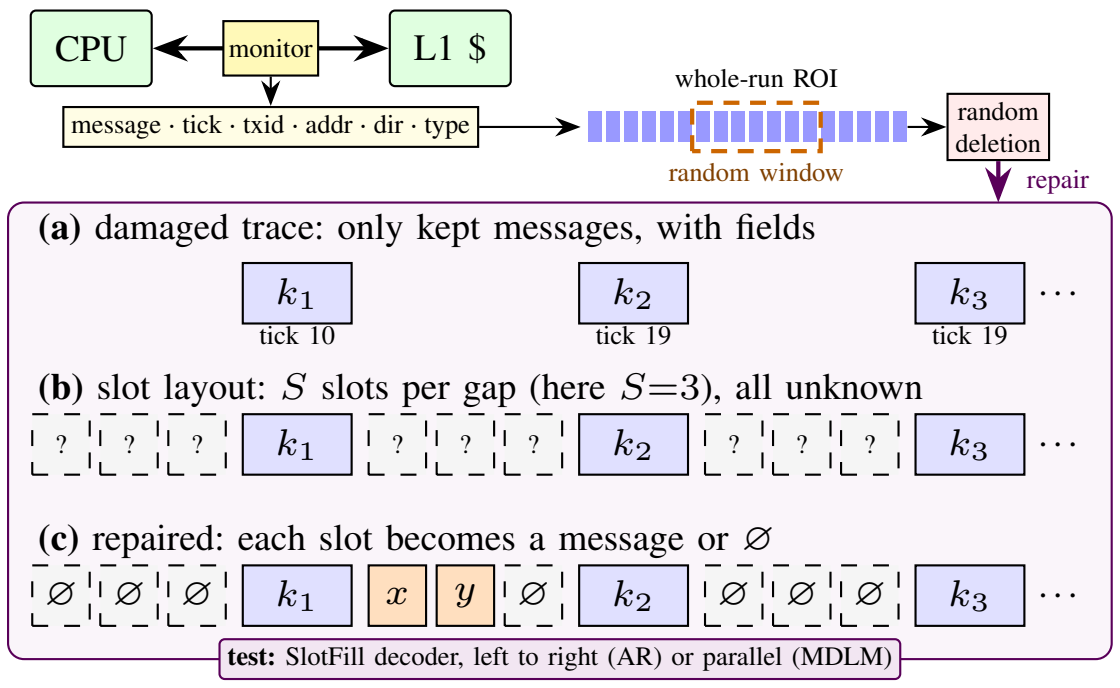

\subsection{Overview of the slot layout}
We place $S$ slots before the first kept message and after every kept message (Fig.~\ref{fig:overview}). The training target puts the deleted messages of a gap into its first slots, in order, and \textsc{null} in the rest. One label per slot thus encodes all three parts of the answer: a non-empty slot says \emph{where} (its gap), its label says \emph{what}, and its index says the \emph{order}. Every gap has the same number of slots, so the input length $S{+}K(S{+}1)$ depends only on $K$ and tells the model nothing about how many messages were deleted. For the same reason we cut windows \emph{after} deletion, as in real use, at $W{=}512$ kept messages; cutting before deletion would let the model infer the loss from how many messages survived. Kept messages are copied, never rewritten, so recovery cannot damage what survived.

\subsection{NULL value and the threshold}
\label{sec:threshold}
Most slots are empty: 97\% at 15\% deletion and 83\% at 50\%. A model that always answers ``empty'' would be right on most slots but would recover nothing, so empty slots never count as correct in any metric (Sec.~\ref{sec:metrics}). Two steps stop the model from drifting to ``empty''. In training, \textsc{null} targets get weight $w{=}0.05$ in the cross-entropy, so a missed message costs 20 times more than a missed empty slot. In decoding, we first add $\log(1/w)$ to the \textsc{null} logit, which undoes this weight and gives back calibrated probabilities. A slot then becomes a message only if $P(\textsc{null})<\tau$. The threshold $\tau$ trades recall for precision: a high $\tau$ inserts more messages, a low $\tau$ fewer. We choose $\tau$ once per model on validation windows (best mean F1 of strict recall and precision over 15, 25 and 50\% deletion) and freeze it for all tests.

\subsection{How many slots}
\label{sec:numslots}
With $S$ slots, a gap that lost $g$ messages can get back at most $\min(g,S)$ of them. So even a perfect model is capped at
\[
C(S) \;=\; \textstyle\sum_{\text{gaps}} \min(g,S) \,\big/\, \sum_{\text{gaps}} g .
\]
$C(S)$ depends only on the gap sizes, and the gap sizes depend only on where the loss events land, not on what the messages are. So we compute $C(S)$ by simulating the deletion process alone (same rate, bursts and window cut) on 3{,}000 windows per rate. The simulated values match $C(S)$ computed from the gap sizes of our test windows within 0.3 points; the latter are the dashed lines in Fig.~\ref{fig:slots}. At 15\% and 25\% deletion, $C(6)\geq 99.7\%$. At 50\%, $C(6)=95.7\%$, and more slots add little: $C(7)=97.5\%$, $C(8)=98.5\%$, $C(9)=99.1\%$, at most 3.4 points. Each slot costs $K{+}1$ positions, so $S{=}9$ would make every window 43\% longer (5{,}129 vs.\ 3{,}590 positions). We use $S{=}6$; Sec.~\ref{sec:slots} tests this choice by also training with $S{=}9$.

\subsection{Filling the slots}
\label{sec:decoders}
The layout works with any model that can label slots. We test two, with the same Transformer~\cite{vaswani2017attention} ($d{=}256$, 6 layers, 8 heads, about 5M parameters), the same features and the same data. Both are small language models trained from scratch on traces only, with no pre-training (Sec.~\ref{sec:size} explains why); the fields, the training recipe and the layout carry most of the work.

\textbf{Left to right (L2R).} The model first reads the damaged window (the kept messages with their fields) with attention in both directions, so it knows what comes after every gap. It then writes the layout one position at a time: each slot sees the whole damaged window and every decision made before it, and outputs a message or \textsc{null}; kept positions are copied. Training uses the true previous slots. Decoding is greedy with a key-value cache: 3{,}590 small steps per window.

\textbf{Parallel (masked diffusion).} The model sees all slots masked and fills them together~\cite{sahoo2024mdlm}. In training, each slot is masked with probability $t\sim U(0.001,0.999)$ and the loss is weighted by $1/t$. Decoding runs 128 steps; each step fixes the most confident slots~\cite{ghazvininejad2019maskpredict,chang2022maskgit} and predicts the rest again with the new context.

\subsection{Monitor fields as features}
\label{sec:features}
Real monitors record more than the message type: AXI carries transaction IDs, trace units add timestamps, and monitor encodings include tags and addresses~\cite{arm2023axi,arm2017coresight,cao2018monitors}. We add four features. Each is computed \textbf{only from the kept messages}, exactly as a debug engineer would see them.
\begin{itemize}[leftmargin=*,nosep]
\item \textbf{Time gap.} The cycles since the previous kept message, bucketed (0--31), as an extra embedding. A long gap hints that something is missing.
\item \textbf{Time-RoPE.} Rotary embeddings~\cite{su2024roformer} rotate attention by position. In half of the heads we rotate by tick instead, so these heads measure distance in cycles. This follows continuous-time encodings for event sequences~\cite{zuo2020thp,kazemi2019time2vec}.
\item \textbf{Orphan flag.} We pair kept requests and responses by txid, preferring the request whose endpoints mirror the response. Each kept message is marked as \emph{partner kept}, \emph{partner not among kept messages}, or \emph{no reply expected}. An unpaired response strongly hints that its request was deleted nearby.
\item \textbf{Address links.} Line addresses are relabelled per window. Each kept message gets the distance (up to 16) to the previous and next kept message with the same address, plus their types.
\end{itemize}

\textbf{Leak audit.} The model only uses what survives deletion; an input built from the clean trace would inflate the results. To check this, we (1) wrote a separate function that builds every model input (tokens, layout and all feature columns) from the damaged window alone, (2) ran it next to our pipeline at deletion rates 0--60\%, and (3) compared the two. The inputs matched exactly in all 160 test windows. For L2R we also compared its two ways of decoding: step by step with a cache, as used in our results, and one pass over the whole layout. Both made exactly the same decisions, so no slot sees a later answer.

\subsection{Metrics}
\label{sec:metrics}
\emph{Strict recall} is the share of deleted messages placed in the right gap, with the right type, at the right position inside the gap. \emph{Precision} is the same count divided by the number of inserted messages. Both count messages only; an empty slot is never counted as a correct answer. We report strict as the main number. Two relaxed metrics explain the errors. \emph{Tie-aware} also accepts a message moved only across identical copies of itself, or across same-tick messages of other flows, since the trace order of same-cycle events between flows is arbitrary. \emph{Flow-relative} accepts the right type anywhere between the nearest kept messages of its own transaction, ignoring order against unrelated flows. In gem5 one txid follows a transaction across all hops; we use this only for scoring, never as model input. We also split results into \emph{CPU$\leftrightarrow$L1} messages and messages \emph{beyond L1} (L2, memory bus, DRAM, walkers, I/O).

\section{Experimental Setup}
\begin{table}[t]
\centering
\caption{gem5 setup and data.}
\label{tab:setup}
\vspace{-3mm}
\footnotesize
\setlength{\tabcolsep}{3pt}
\begin{tabular}{@{}p{2.0cm}p{6.1cm}@{}}
\toprule
CPU / ISA & TimingSimpleCPU, x86; 4 and 8 cores \\
Caches & private L1I/L1D per core, shared L2 behind a snooping L2 bus; classic memory system \\
Memory & DDR4 (SE), DDR3 (FS) \\
Monitors & CommMonitor on all major links, recording requests and responses. Per core (6): cpu--icache, cpu--dcache, icache--l2bus, dcache--l2bus, iwalker--l2bus, dwalker--l2bus. Shared (3): l2bus--l2cache, l2cache--membus, membus--dram. FS only (18, OS, interrupt, PIO and DMA traffic): membus--bridge and bridge--iobus (both directions), iobus to cmos, dma1, keyboard, pic1, pic2, pit, speaker, io\_apic, uart; io\_apic--iobus; pcibus--ide (both directions); iobus--iocache, iocache--membus \\
Monitor count & 27 / 51 (SE-4/8), 45 / 69 (FS-4/8) \\
Workloads & SE: fft, lu, radix, snoop (Peterson lock), threads, ocean. FS (Ubuntu, after-boot checkpoint): blackscholes, fluidanimate, streamcluster, io, snoop, threads, swaptions \\
Unseen program & ocean (SE), swaptions (FS): never in training \\
Training data & 104 train / 13 val traces per set, $\approx$50k messages each; every 2nd train trace of 4c and 8c, one model per mode ($\approx$5.4M messages) \\
Test data & whole ROI, no pause, no filter; 5 random 100k-message segments per run \\
Message types & 407 (FS train), 322 (SE train) \\
\bottomrule
\end{tabular}
\end{table}

\textbf{Traces.} Table~\ref{tab:setup} summarizes the gem5 setup. It uses gem5's classic memory system with 4 and 8 busy cores. Messages from all monitors are merged in tick order. Workloads come from PARSEC~\cite{bienia2008parsec}, SPLASH-2~\cite{woo1995splash2}, and two small multithreaded microbenchmarks (threads; snoop, a Peterson lock). Traces are split \emph{by run}, so no run appears in both training and test. One program per mode is held out completely for the unseen program test.

\textbf{Natural test traces.} The training corpus was recorded in windows of about 50k messages; the simulator paused and finished pending transactions between windows. Real chips never pause, so we test only on \emph{whole-run} traces. For each program and core count, gem5 records the whole region of interest as one uninterrupted stream, with no pausing and no filtering. We used the smallest standard inputs (PARSEC \texttt{test}, \texttt{simsmall} for blackscholes). The 26 runs contain 2.07B messages. We take 5 segments of 100k messages per run at random positions, chosen by position only, never by content. The segments cover 99.3--100\% of each run's messages by type.

\textbf{Training.} Windows hold 512 kept messages and $S{=}6$ slots per gap; deletion rates are drawn per window from $U(0,0.6)$, in bursts of 1--2 that can merge into longer gaps. All models are trained with the same recipe (learning rate $10^{-3}$), the same number of epochs for both decoders within each experiment type, and keep the checkpoint with the lowest validation loss. Retraining L2R on FS with two more seeds varies strict recovery by at most 0.8 points (std) at 15--25\% and 1.9 at 50\% (FS-8: 80.7$\pm$1.9\%), so results are consistent across seeds; the weakest seed still beats the parallel decoder in every FS setting. Thresholds chosen on validation: $\tau{=}0.5$ for L2R (FS and SE), 0.6 (FS) and 0.5 (SE) for the parallel decoder. We test at 15\%, 25\% and 50\% deletion, on 128 windows of 512 kept messages per program and rate, drawn from these segments: 768--896 windows and about 69k (15\%) to 460k (50\%) deleted messages per set and rate.

\textbf{Is the data enough?} About 5.4M training messages per mode cover the message vocabulary: over 99\% of each shard's message types are in the training vocabulary, types missing from one shard appear in another, and fewer than 0.04\% of the FS test messages are of a type never seen in training. Validation loss barely changed in the last epochs. More such traces would add little, so we study model size instead (Sec.~\ref{sec:size}).

\section{Results}
\textbf{Guide.} Tables~\ref{tab:main} and~\ref{tab:groups} compare both decoders (L2R and parallel) in both modes (FS and SE). Tables~\ref{tab:s9} and~\ref{tab:ablation} (slots, model size) use L2R on FS. Table~\ref{tab:fields} (field ablation, coarse time, other loss patterns) uses both decoders on FS, the harder mode with more message types. Tables~\ref{tab:automodel} and~\ref{tab:o3} (debugging tests, out-of-order CPU) use L2R on SE, since AutoModel flow models exist only for the SE programs and we performed a brief O3 CPU test in SE mode only.

\begin{table*}[t]
\centering
\caption{Recovery on whole-run gem5 traces (\% of deleted messages; strict precision in brackets). Strict = right gap, type and order. L2R = left to right; Par.\ = parallel (masked diffusion). Same layout, model size, features and data. Bigram$^*$ is given the true gap positions and sizes. Txid rule inserts the missing partner of every kept request or reply next to it. Prior: the find-then-fill of~\cite{fender2024} with a BERT masked LM, reimplemented at our model size, data and training budget; no monitor fields, at most one message per gap.}
\label{tab:main}
\vspace{-3mm}
\small
\setlength{\tabcolsep}{4pt}
\begin{tabular}{@{}ll ccc c ccc c ccc@{}}
\toprule
 & & \multicolumn{3}{c}{15\% deleted} && \multicolumn{3}{c}{25\% deleted} && \multicolumn{3}{c}{50\% deleted} \\
\cmidrule{3-5}\cmidrule{7-9}\cmidrule{11-13}
Set & Dec. & Strict (prec.) & Tie-aware & Flow-rel. && Strict (prec.) & Tie-aware & Flow-rel. && Strict (prec.) & Tie-aware & Flow-rel. \\
\midrule
\multirow{5}{*}{FS, 4 cores} & L2R & \textbf{94.1} (91.3) & 95.1 & 95.4 && \textbf{92.5} (90.7) & 93.9 & 94.3 && \textbf{82.1} (84.4) & 85.5 & 86.6 \\
 & Par. & 91.7 (92.1) & 92.7 & 92.8 && 89.6 (90.9) & 91.0 & 91.3 && 78.2 (84.3) & 81.5 & 82.4 \\
 & Bigram$^*$ & 57.1 (57.1) & 58.0 & 59.2 && 52.9 (53.1) & 54.6 & 56.7 && 40.3 (42.2) & 45.0 & 49.7 \\
 & Prior & 19.6 (21.0) & 31.8 & 35.3 && 18.0 (22.8) & 30.6 & 34.5 && 12.3 (24.7) & 23.6 & 27.7 \\
 & Txid rule & 3.3 (9.8) & 4.3 & 18.4 && 2.5 (11.3) & 3.5 & 14.0 && 1.3 (13.5) & 2.5 & 7.7 \\
\midrule
\multirow{5}{*}{FS, 8 cores} & L2R & \textbf{93.7} (90.1) & 94.8 & 95.1 && \textbf{92.2} (89.9) & 93.7 & 94.1 && \textbf{81.9} (83.9) & 85.3 & 86.5 \\
 & Par. & 91.1 (91.2) & 92.1 & 92.3 && 89.1 (90.2) & 90.5 & 90.8 && 77.2 (83.4) & 80.6 & 81.6 \\
 & Bigram$^*$ & 53.2 (53.2) & 54.0 & 55.4 && 49.4 (49.6) & 50.9 & 53.4 && 36.9 (38.6) & 41.0 & 46.7 \\
 & Prior & 18.9 (17.8) & 31.6 & 34.9 && 17.6 (20.2) & 30.6 & 34.2 && 11.8 (22.4) & 23.0 & 27.1 \\
 & Txid rule & 2.7 (6.4) & 3.5 & 21.6 && 2.0 (7.7) & 2.9 & 16.4 && 1.1 (9.3) & 2.1 & 8.9 \\
\midrule
\multirow{5}{*}{SE, 4 cores} & L2R & \textbf{96.8} (94.6) & 97.3 & 97.3 && \textbf{95.4} (93.4) & 96.3 & 96.4 && \textbf{86.5} (87.9) & 89.3 & 89.9 \\
 & Par. & 95.2 (95.3) & 95.7 & 95.8 && 93.4 (94.0) & 94.3 & 94.4 && 82.7 (88.1) & 85.7 & 86.4 \\
 & Bigram$^*$ & 58.1 (58.1) & 59.7 & 60.7 && 54.2 (54.4) & 57.3 & 58.9 && 41.4 (43.4) & 49.4 & 53.5 \\
 & Prior & 21.5 (19.8) & 32.5 & 34.8 && 19.2 (22.0) & 30.6 & 33.2 && 12.7 (24.7) & 22.8 & 25.7 \\
 & Txid rule & 2.9 (9.7) & 3.7 & 18.2 && 2.1 (10.7) & 3.1 & 13.5 && 1.2 (13.1) & 2.2 & 7.2 \\
\midrule
\multirow{5}{*}{SE, 8 cores} & L2R & \textbf{93.5} (88.6) & 94.7 & 94.9 && \textbf{91.6} (87.1) & 93.4 & 93.6 && \textbf{78.6} (76.4) & 84.4 & 85.7 \\
 & Par. & 91.9 (91.0) & 92.9 & 93.0 && 89.4 (90.0) & 90.9 & 91.2 && 75.9 (82.4) & 80.5 & 81.8 \\
 & Bigram$^*$ & 51.7 (51.7) & 52.8 & 54.3 && 47.5 (47.6) & 49.9 & 52.5 && 34.9 (36.6) & 41.5 & 47.6 \\
 & Prior & 20.4 (16.6) & 32.6 & 35.3 && 18.1 (18.9) & 30.2 & 33.5 && 11.7 (21.6) & 21.9 & 25.5 \\
 & Txid rule & 1.8 (3.9) & 2.5 & 23.9 && 1.3 (4.7) & 2.1 & 18.0 && 0.8 (6.2) & 1.6 & 9.4 \\
\bottomrule
\end{tabular}
\end{table*}

\subsection{Main results}
Table~\ref{tab:main} shows recovery on the whole-run traces. With either decoder, at 15\% deletion SlotFill puts over 90\% of deleted messages back exactly: right gap, right type, right order. Precision is about the same as recall, so it rarely inserts messages that were never there. At 50\% deletion, where half of all messages are gone, it still restores 76--83\% exactly with the parallel decoder and 79--87\% with L2R.

\textbf{Simple baselines.} A bigram model of the training traces that is even told the true gap positions and sizes restores 52--58\% at 15\% and 35--41\% at 50\%, 37--45 points below L2R in every set. A txid rule that inserts the learned partner of every orphan next to it restores under 4\% exactly (up to 24\% flow-relative): the orphan flag tells which message is missing, but not where.

\textbf{Prior work.} Fender's thesis~\cite{fender2024} gives no model size, data size or training length, so we reimplemented its find-then-fill with a BERT masked LM at our model size (larger models add nothing, Table~\ref{tab:ablation}), data and training. It restores 19--22\% at 15\% and 12--13\% at 50\% (precision 17--25\%), far below SlotFill even without fields (78--83\% at 15\%, Table~\ref{tab:fields}). Neither uses monitor fields, so this gap comes from our model, training and output format: Fender finds at most one message per gap, only where BERT mispredicts a kept message, while SlotFill decides where, what and in which order in one pass.

\textbf{Left to right vs.\ parallel.} The L2R decoder recovers more in all 12 settings (4 sets $\times$ 3 rates): 1.6--3.1 points at 15--25\% deletion and 2.7--4.7 points at 50\%. Its precision is within about 1 point on FS and SE-4, but on SE-8 it inserts more messages that were never there (precision 2--6 points lower, 76.4\% vs.\ 82.4\% at 50\%). The largest gain is beyond L1: 4--12 points (Table~\ref{tab:groups}). We think the reason is how decisions are made. L2R makes about 3{,}000 slot decisions one at a time, and each one sees all earlier decisions. The parallel decoder fixes about 24 slots per step, and slots fixed in the same step cannot see each other, which hurts most in long gaps and in tangled traffic beyond L1. L2R needs more steps, but each step is small: on one NVIDIA A100 (80\,GB) it repairs a 512-message window in about 0.45\,s, against 1.1\,s for the parallel decoder, which runs the full model over the whole layout in each of its 128 steps.

\textbf{Error types.} For the parallel decoder the relaxed metrics add only about 1 point at 15\% and 3--6 at 50\%: deleted messages are mostly restored exactly or missed. Eight cores are harder than four in SE (91.9 vs.\ 95.2\% at 15\%, more interleaving); in FS the two are close.

\begin{table}[t]
\centering
\caption{Strict recovery by message group and on the unseen program (whole-run test). L2R = left to right; Par.\ = parallel.}
\label{tab:groups}
\vspace{-3mm}
\footnotesize
\setlength{\tabcolsep}{2.6pt}
\begin{tabular}{@{}ll cc cc ccc@{}}
\toprule
 & & \multicolumn{2}{c}{CPU$\leftrightarrow$L1} & \multicolumn{2}{c}{Beyond L1} & \multicolumn{3}{c}{Unseen program} \\
\cmidrule(lr){3-4}\cmidrule(lr){5-6}\cmidrule(l){7-9}
Set & Dec. & 15\% & 50\% & 15\% & 50\% & 15\% & 25\% & 50\% \\
\midrule
\multirow{2}{*}{FS-4} & L2R & 95.9 & 89.0 & 82.6 & 65.4 & 92.7 & 90.9 & 80.6 \\
 & Par. & 94.2 & 85.9 & 75.3 & 55.1 & 90.3 & 88.0 & 77.2 \\
\multirow{2}{*}{FS-8} & L2R & 96.0 & 88.9 & 78.0 & 63.3 & 91.5 & 90.6 & 82.2 \\
 & Par. & 94.2 & 84.8 & 69.4 & 51.7 & 88.7 & 86.6 & 77.4 \\
\multirow{2}{*}{SE-4} & L2R & 97.0 & 90.9 & 94.9 & 87.1 & 93.5 & 90.5 & 79.3 \\
 & Par. & 95.7 & 87.5 & 90.8 & 77.1 & 93.3 & 90.3 & 78.7 \\
\multirow{2}{*}{SE-8} & L2R & 94.4 & 83.0 & 85.7 & 76.2 & 90.0 & 86.8 & 72.0 \\
 & Par. & 93.6 & 81.1 & 78.1 & 64.7 & 89.9 & 86.4 & 73.2 \\
\bottomrule
\end{tabular}
\vspace{1mm}

{\footnotesize Beyond-L1 share of deleted messages: 12--13\% (FS), 8--10\% (SE). Unseen program: swaptions (FS), ocean (SE).}
\end{table}

\subsection{Where the errors are}
Table~\ref{tab:groups} splits the results. Messages between a CPU and its L1 follow program order and are easy: 94--97\% at 15\% for both decoders (93.6\% for parallel SE-8). Messages beyond L1 (cache misses, evictions, snoops, walkers, I/O) depend on cache state and on other cores, and are harder: 69--91\% at 15\% for the parallel decoder and 78--95\% for L2R. The FS \texttt{io} program, with 30--35\% of its deletions beyond L1, is the hardest (parallel, 8 cores: 84.9\% at 15\%, 65.5\% at 50\%).

\textbf{Unseen programs.} The held-out programs were never in training. At low loss they score close to the overall results (Table~\ref{tab:groups} vs.\ Table~\ref{tab:main}): at 15\% at most 2.4 points lower for the parallel decoder and at most 3.5 points lower for L2R. On FS, L2R stays within 2.2 points at every rate. At 50\% the gap grows for SE's \texttt{ocean} (up to 7 points with L2R, 4 with the parallel decoder), so a new program costs more when much of the trace is lost.

\subsection{How many slots are needed}
\label{sec:slots}
\begin{figure}[t]
\centering
\includegraphics[width=\columnwidth]{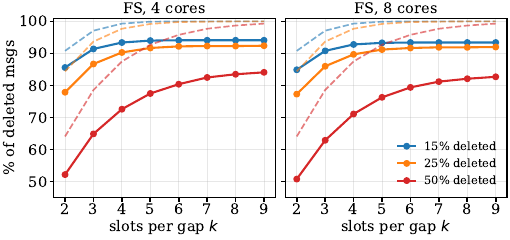}
\vspace{-6mm}
\caption{Effect of the slot limit (L2R trained with $S{=}9$, FS whole-run test, 32 windows per program and rate). Solid: strict recovery. Dashed: the cap $C(k)$ of Sec.~\ref{sec:numslots}, the most any $k$-slot model could recover.}
\label{fig:slots}
\end{figure}

\begin{table}[t]
\centering
\caption{L2R trained with $S{=}6$ vs.\ $S{=}9$ slots per gap (FS whole-run, strict \%, 32 windows per program and rate). More slots help only at 50\% loss, and precision drops.}
\label{tab:s9}
\vspace{-3mm}
\footnotesize
\setlength{\tabcolsep}{4.5pt}
\begin{tabular}{@{}ll ccc c@{}}
\toprule
Set & $S$ & 15\% & 25\% & 50\% & Prec.\ at 50\% \\
\midrule
\multirow{2}{*}{FS-4} & 6 & \textbf{94.4} & \textbf{92.7} & 82.5 & \textbf{84.6} \\
 & 9 & 94.1 & 92.4 & \textbf{84.1} & 83.9 \\
\multirow{2}{*}{FS-8} & 6 & \textbf{94.2} & \textbf{92.6} & 82.2 & \textbf{84.0} \\
 & 9 & 93.4 & 92.0 & \textbf{82.7} & 81.3 \\
\midrule
\multicolumn{6}{@{}l}{Window length: 3{,}590 ($S{=}6$) vs.\ 5{,}129 ($S{=}9$) positions (+43\%)} \\
\bottomrule
\end{tabular}
\end{table}

Loss events land at random and merge, so gaps grow with the rate: at 50\% deletion, 36\% of deleted messages lie past the 2nd place of their gap. To see how many slots are used, we lock slots $k{+}1$ to $S$ to \textsc{null} and decode with at most $k$ messages per gap. With the parallel decoder ($S{=}6$), strict recovery at 50\% rises from 47--49\% at $k{=}2$ to 78\% at $k{=}6$, while at 15\% it is flat after 4 slots. To test $S{=}6$ itself, we also trained L2R with $S{=}9$ (Fig.~\ref{fig:slots}). At 15--25\% loss its curves are flat after 5 slots; at 50\% they still rise slowly from $k{=}6$ to 9 (80.4 to 84.1\% on 4 cores, 79.4 to 82.7\% on 8 cores), as the cap allows. Against the $S{=}6$ L2R model (Table~\ref{tab:s9}), the $S{=}9$ model gains only 0.5--1.6 points at 50\% loss, is 0.3--0.8 points lower at 15--25\%, has lower precision at 50\% (by 0.7 and 2.7 points), and makes every window 43\% longer, raising training and decoding cost. Even with 9 slots, the model stays 15--17 points below the cap at 50\%, so better placement, not more slots, is where the remaining gain lies. Since the loss rate is unknown when a trace is repaired, one $S$ must serve all rates. $S{=}6$ is enough at typical loss; where heavy loss is expected, $S$ can simply be raised at training time.

\subsection{Why a small model}
\label{sec:size}
\label{sec:base}
\begin{table}[t]
\centering
\caption{Model size (FS whole-run, strict \%). Same data, features and epochs.}
\label{tab:ablation}
\vspace{-3mm}
\footnotesize
\setlength{\tabcolsep}{4pt}
\begin{tabular}{@{}l cc cc@{}}
\toprule
 & \multicolumn{2}{c}{4 cores} & \multicolumn{2}{c}{8 cores} \\
\cmidrule(lr){2-3}\cmidrule(l){4-5}
Model & 15\% & 50\% & 15\% & 50\% \\
\midrule
L2R, 2.2M & 93.0 & 79.8 & 92.4 & 79.1 \\
L2R, 5.3M (ours) & 94.1 & 82.1 & 93.7 & 81.9 \\
L2R, 9.7M & 94.1 & 82.1 & 93.8 & 81.8 \\
\midrule
Parallel, 5.0M & 91.7 & 78.2 & 91.1 & 77.2 \\
\bottomrule
\end{tabular}
\end{table}

We train small models (5M parameters) from scratch instead of adapting large pretrained language models, for two reasons. First, trace messages are not natural language. Each message type is one token for our model, but a general tokenizer splits it into about 13 sub-word tokens, so a 512-message window becomes thousands of tokens, and knowledge learned from text does not carry over to cache and memory traffic. Second, adapting a 7--8B model by LoRA or full fine-tuning keeps a large, costly model whose pre-training does not fit the task. Table~\ref{tab:ablation} shows how size matters. Halving the L2R model to 2.2M parameters costs only 1--1.5 points at 15\% loss and 2--3 points at 50\%, makes decoding about 30\% faster; even the 2.2M L2R model beats the 5M parallel one at every rate. Doubling it to 9.7M changes nothing (within 0.1 points at every rate) and makes decoding about 30\% slower. Accuracy has levelled off by 5M, so we use 5M.

\subsection{Fields, coarse time and other loss patterns}
\label{sec:ablation}
\label{sec:robust}
\begin{table}[t]
\centering
\caption{FS whole-run test, strict \%. No fields: retrained without time gap, time-RoPE and orphan flag. Mixed: retrained with ticks rounded to 1, 10, 100 or 1000 cycles at random, and with loss drawn per window from bursts of 1--2 (50\%), bursts of 3--6 (25\%) or a FIFO (25\%); tested on 32 windows per program and rate. $C(6)$: share of deleted messages that fit in 6 slots. Rounding to 100 cycles gives the same results as 1000 (within 0.5 points). $^\dagger$Real loss 23\% and 28\%: see text.}
\label{tab:fields}
\label{tab:robust}
\vspace{-3mm}
\footnotesize
\setlength{\tabcolsep}{2.6pt}
\begin{tabular}{@{}ll ccc ccc@{}}
\toprule
 & & \multicolumn{3}{c}{FS-4} & \multicolumn{3}{c}{FS-8} \\
\cmidrule(lr){3-5}\cmidrule(l){6-8}
Model & Test & 15\% & 25\% & 50\% & 15\% & 25\% & 50\% \\
\midrule
Full, L2R & exact & 94.1 & 92.5 & 82.1 & 93.7 & 92.2 & 81.9 \\
Full, Par. & exact & 91.7 & 89.6 & 78.2 & 91.1 & 89.1 & 77.2 \\
No fields, L2R & -- & 80.8 & 77.0 & 58.4 & 82.5 & 79.4 & 62.7 \\
No fields, Par. & -- & 78.5 & 74.0 & 55.2 & 80.7 & 76.7 & 60.1 \\
\midrule
\multirow{5}{*}{\shortstack[l]{Mixed,\\L2R}} & exact & 92.7 & 90.5 & 77.5 & 92.3 & 90.2 & 77.6 \\
 & ticks /10 & 80.5 & 75.8 & 58.9 & 82.5 & 79.2 & 62.4 \\
 & ticks /1000 & 79.1 & 74.2 & 57.3 & 81.8 & 78.1 & 61.2 \\
 & bursts 3--6 & 82.7 & 77.6 & 55.7 & 82.3 & 77.8 & 56.3 \\
 & FIFO & 93.4 & 93.5 & 89.5 & 93.2$^\dagger$ & 92.3$^\dagger$ & 79.0 \\
\midrule
\multirow{5}{*}{\shortstack[l]{Mixed,\\Par.}} & exact & 88.1 & 86.0 & 73.5 & 87.6 & 85.7 & 72.7 \\
 & ticks /10 & 77.0 & 72.6 & 54.8 & 78.9 & 75.3 & 59.0 \\
 & ticks /1000 & 75.8 & 70.9 & 52.7 & 77.8 & 74.2 & 57.6 \\
 & bursts 3--6 & 72.3 & 69.2 & 50.4 & 71.6 & 68.3 & 50.0 \\
 & FIFO & 89.8 & 90.3 & 84.8 & 91.2$^\dagger$ & 89.4$^\dagger$ & 72.8 \\
\midrule
\multicolumn{2}{@{}l}{$C(6)$, bursts 3--6} & 94.5 & 91.2 & 74.2 & 94.5 & 91.4 & 74.2 \\
\bottomrule
\end{tabular}
\end{table}

\textbf{Monitor fields.} Exact global cycle counts are optimistic for silicon, where trace timestamps are often coarse or sent only periodically~\cite{arm2017coresight}. We therefore retrain without the time features (time gap and time-RoPE) and without the orphan flag, on the same data, for both decoders on FS, and test on the same whole-run traces with the same thresholds (Table~\ref{tab:fields}). The ablation models were trained longer than the full models, until their loss curve flattened, so the measured gaps are a lower bound.

Without these fields, strict recovery drops by 10--13 points at 15\% deletion and by 17--24 points at 50\%, for both decoders. Precision drops too (parallel, FS-4 at 50\%: 68.0\% vs.\ 84.3\%), so the model inserts more messages that were never there. Without the fields, L2R keeps a 2--3 point lead. The loss is largest on the unseen program: at 15\%, swaptions falls from 90.3\% to 75.1\% (parallel, 4 cores) and from 88.7\% to 66.7\% (parallel, 8 cores). The fields therefore help the model generalize, not only fit. Even without them, both decoders still restore 78--83\% at 15\% deletion, so SlotFill stays useful on traces without timestamps or transaction IDs. We removed both fields together, so this test does not separate the time features from the orphan flag.

\textbf{Coarse time.} A model trained on exact cycles does not transfer to rounded ticks: shown ticks rounded to 10 or 1000 cycles, the full L2R model recovers only 36--49\% at 15\%, less than the model with no time at all, since most messages are 0--1 cycles apart and a rounded gap of 0 reads as ``nothing missing''. We therefore retrain both decoders \emph{mixed}: each training window sees its ticks rounded to 1, 10, 100 or 1000 cycles, at random, and one of three loss patterns (caption of Table~\ref{tab:fields}). One model then serves every resolution. On rounded ticks it recovers 76--83\% at 15\% and 53--62\% at 50\%, about the same as the model without time features, so rounded time adds little beyond the other fields. On exact ticks L2R gives up about 2 points at 15\% and 5 points at 50\% against the exact-time model on the same 32 windows (Table~\ref{tab:s9}). A monitor with exact time should use a model trained on exact time; when the resolution is coarse or unknown, the mixed model is a safe default. L2R stays ahead of the parallel decoder at every resolution, by 3.3--4.7 points at 15\%.

\textbf{Other loss patterns.} The mixed models also train on longer bursts and FIFO loss. \emph{Longer bursts}: bursts of 3--6 messages make longer gaps, so only 94.5\% (15\%) and 74.2\% (50\%) of the deleted messages fit in 6 slots. L2R stays 12 points below this cap at 15\% and 18--19 points at 50\%; the parallel decoder falls 10 points further behind at 15\%. \emph{Loss tied to traffic}: a 16-entry trace FIFO that drains at a fixed rate drops every message that arrives while it is full, so loss lands in dense traffic; the drain rate is set per window to reach the target rate. On 4 cores, both decoders do better than under random loss (L2R 89.5\%, parallel 84.8\% at 50\%), since FIFO loss hits regular, dense traffic. On 8 cores, more than 16 messages sometimes arrive in one cycle, so this FIFO cannot lose less than 23\%. At 50\%, the loss rate is right but the FIFO drops long runs of messages, so many gaps exceed 6 slots and only 87\% of the lost messages can be restored.

\subsection{Does repair help debugging?}
\label{sec:automodel}
\begin{table}[t]
\centering
\caption{Debugging on SE traces (L2R repair). AutoModel~\cite{ahmed2024automodel} flow models mined from other clean runs (pooled, 5 traces $\times$ 4 models). Top: relative acceptance RA = accepted events / accepted events in the clean trace (repaired: kept messages plus inserted ones that match a message lost in their gap). Bottom: 480--500 injected bugs per type and setting; relative rejection RR = rejected bugs / rejected without loss; the checker itself rejects only 6.4\% of dropped replies, so their RR is noisy.}
\label{tab:automodel}
\vspace{-3mm}
\footnotesize
\setlength{\tabcolsep}{3.2pt}
\begin{tabular}{@{}l c ccc ccc@{}}
\toprule
 & Acceptance & \multicolumn{3}{c}{Damaged, RA (\%)} & \multicolumn{3}{c}{Repaired, RA (\%)} \\
\cmidrule(lr){3-5}\cmidrule(l){6-8}
Program & on clean (\%) & 15\% & 25\% & 50\% & 15\% & 25\% & 50\% \\
\midrule
threads, 4c & 71.2 & 84.8 & 74.7 & 49.7 & 99.8 & 99.5 & 95.4 \\
threads, 8c & 83.0 & 84.6 & 74.6 & 49.5 & 99.8 & 99.5 & 95.9 \\
snoop, 4c & 85.3 & 84.8 & 74.7 & 49.7 & 99.9 & 99.7 & 97.1 \\
snoop, 8c & 75.2 & 84.7 & 74.6 & 49.5 & 99.4 & 98.8 & 93.3 \\
\midrule
 & Rejection & \multicolumn{3}{c}{Damaged, RR (\%)} & \multicolumn{3}{c}{Repaired, RR (\%)} \\
\cmidrule(lr){3-5}\cmidrule(l){6-8}
Bug type & on buggy (\%) & 15\% & 25\% & 50\% & 15\% & 25\% & 50\% \\
\midrule
wrong type    & 75.2 & 82.8 & 74.7 & 46.9 & 82.8 & 74.3 & 45.2 \\
spurious      & 39.2 & 52.0 & 41.1 & 21.9 & 51.8 & 38.8 & 17.6 \\
swapped order & 40.7 & 45.2 & 34.6 & 22.6 & 44.5 & 31.2 & 21.6 \\
dropped reply & 6.4 & 75.0 & 65.6 & 40.6 & 85.9 & 68.8 & 43.7 \\
\midrule
bug re-inserted (\%) & -- & -- & -- & -- & 90.6 & 87.6 & 79.6 \\
\bottomrule
\end{tabular}
\end{table}

Checks on a repaired trace should match those on a complete one. We check each SE run of \texttt{threads} and \texttt{snoop} against flow models that AutoModel~\cite{ahmed2024automodel} mined from four other clean runs of the same program (Table~\ref{tab:automodel}, top). The relative acceptance (RA) of the damaged trace falls to 49.5--84.8\%, mainly because lost events are absent. Repair raises it to 98.8--99.9\% at 15--25\% loss and 93.3--97.1\% at 50\%; accepted wrong insertions add 0.1--1.6\% at 15--25\% and up to 7.4\% at 50\%. What loss really does to the checker is false alarms: loss makes the checker reject correct surviving messages (about 475 per 100k-message trace at 50\% loss, 8-core \texttt{threads}); after repair about 40 remain. Mining models \emph{from} repaired traces is riskier, as wrong insertions add false transitions.

\textbf{Does repair hide bugs?} SlotFill is trained only on bug-free traces, so we test whether repair makes bugs harder to reject. We inject four bug types into each clean SE run, about one per 1{,}024 messages (a dropped reply, a wrong command, a duplicated reply, a reply moved before its request), delete 15--50\%, repair, and count the bugs the same models reject (Table~\ref{tab:automodel}, bottom). Repair lowers the relative rejection (RR) by at most 4.3 points against the damaged trace: kept messages never change, but a few inserted ones make a kept bug fit its context. Loss costs far more, since a deleted bug cannot be seen at all (wrong commands: 75\% rejected without loss, RR 47\% at 50\%). The one risk is a dropped reply, which looks like a lost one, so repair puts it back in 80--91\% of cases; inserted messages are marked, so checks for missing replies must use kept messages only. Each inserted message also carries the model's confidence: on the test traces, a threshold of 0.9 marks 37\% of wrongly inserted replies but only 2.4\% of correct ones, so an engineer can check the uncertain ones first.

\textbf{Use case of this work.} In post-silicon validation, a trace buffer records on-chip traffic during a test, and the trace is checked against expected flows. When the buffer overflows, lost messages make correct behavior look wrong, and engineers chase alarms that are not bugs. SlotFill runs as a cleanup step before the check: it repairs broken flows, removes most of these false alarms, never changes a recorded message and marks every inserted message.

\subsection{An out-of-order CPU}
\label{sec:o3}
\begin{table}[t]
\centering
\caption{SE traces from an out-of-order (O3) CPU, L2R, strict \%. Tie-aware counts messages in the same cycle as interchangeable. In-order: Table~\ref{tab:main}.}
\label{tab:o3}
\vspace{-3mm}
\footnotesize
\setlength{\tabcolsep}{3pt}
\begin{tabular}{@{}l ccc ccc@{}}
\toprule
 & \multicolumn{3}{c}{4 cores} & \multicolumn{3}{c}{8 cores} \\
\cmidrule(lr){2-4}\cmidrule(l){5-7}
Train $\rightarrow$ test & 15\% & 25\% & 50\% & 15\% & 25\% & 50\% \\
\midrule
In-order $\rightarrow$ in-order & 96.8 & 95.4 & 86.5 & 93.5 & 91.6 & 78.6 \\
In-order $\rightarrow$ O3 & 14.0 & 13.0 & 9.8 & 14.8 & 13.6 & 10.5 \\
O3 $\rightarrow$ O3 & 54.5 & 51.1 & 39.9 & 45.3 & 42.1 & 32.6 \\
\quad tie-aware & 71.4 & 70.6 & 62.5 & 62.9 & 62.2 & 56.1 \\
\bottomrule
\end{tabular}
\end{table}

The O3 test uses the same programs, windows and scoring as the in-order SE test. A model trained on in-order traces does not transfer: it recovers 14--15\% at 15\% loss and inserts many false messages. Trained on O3 traces instead (same model and epochs and a similar amount of training data, threshold $\tau{=}0.2$ chosen on O3 validation), it recovers 45--55\% at 15\% and 33--40\% at 50\%, with precision close to recall. Many remaining errors are in the order of messages that share a cycle, which an O3 core produces far more often: counting these as interchangeable, it reaches 63--71\% at 15\%. Recovery varies by program, from 91\% on \texttt{threads} to 16--30\% on \texttt{fft} and \texttt{ocean}.

\section{Limitations}
Our traces come from gem5, not silicon, mostly with in-order CPUs, and bugs are injected after simulation. Training covers random bursts and FIFO loss only, and a buffer that drops hundreds of messages in a row cannot be rebuilt. A rigorous study of insertion models~\cite{stern2019insertion,gu2019levenshtein,shen2020blank}, training on buggy traces, large language models, and O3 in both modes are left to future work.

\section{Conclusion}
SlotFill recovers deleted messages in SoC communication traces without placeholders: it decides where each message was, what it was, and in which order. One model does all three, using a slot layout. Every gap gets the same number of slots, so the input hides how much was lost, one label per slot answers all three, and kept messages are never rewritten. The layout works with different decoders; filling slots left to right works best. On whole-run gem5 traces from 4- and 8-core systems, SlotFill restores over 90\% of deleted messages exactly at 15\% deletion and over three quarters at 50\%, and stays close on programs never seen in training. Repaired traces pass a flow checker like complete ones and barely change bug rejection, so SlotFill can serve as a cleanup step before post-silicon checks.

\smallskip
\noindent\textbf{Acknowledgement.}
This work is supported by a National Science Foundation Grant No. 2434247. 
Any opinions, findings, and conclusions 
expressed in this material are those of the author(s) and do not necessarily reflect the views of the funding agencies.

\smallskip
\noindent\textbf{Content Generated by AI.}
An LLM assistant was used to help draft and edit the text of the paper. All experiments were designed, run and checked by the authors, who take responsibility for all content.

\clearpage
\bibliographystyle{IEEEtran}
\bibliography{refs}

\end{document}